\documentclass[showpacs,preprintnumbers,amsmath,amssymb]{revtex4-1}
\usepackage{epsfig}
\usepackage{epsfig,amsmath}
\usepackage{graphicx}
\usepackage{dcolumn}

\usepackage{stmaryrd}
\usepackage{mathrsfs}
\usepackage{pifont}
\usepackage{amsthm}
\usepackage{amssymb}
\usepackage{amsmath}
\usepackage{epsfig,amsmath}
\usepackage{graphicx}
\usepackage{dcolumn}
\usepackage{bm}

\newcommand{\beq}{\begin{equation}}
\newcommand{\eeq}{\end{equation}}
\newcommand{\beqa}{\begin{eqnarray}}
\newcommand{\eeqa}{\end{eqnarray}}

\newcommand{\da}{\dagger} 
 \newcommand{\si}{\sigma}

\newcommand{\la}{\langle}
\newcommand{\ra}{\rangle}
\newcommand{\pr}{\prime}
\newcommand{\non}{\nonumber}

\def\pra#1{{ Phys.\ Rev. A\/} {\bf#1}}
\def\prb#1{{ Phys.\ Rev. B\/} {\bf#1}}
\def\pre#1{{ Phys.\ Rev. E\/} {\bf#1}}
\def\prl#1{{ Phys.\ Rev.\ Lett.} {\bf#1}}

\def\rmp#1{{ Rev. \ Mod. \ Phys.} {\bf#1}}

\begin{document}

\title{Entanglement Dephasing Dynamics Driven by a Bath of Spins}

\author{Jie Xu\footnote{Email address:
jxu2@stevens.edu}, Jun Jing and Ting Yu\footnote{Email
address: ting.yu@stevens.edu}}

\affiliation{Center for Controlled Quantum Systems, and the
Department of Physics and Engineering Physics, Stevens Institute of
Technology, Hoboken, New Jersey 07030, USA}


\date{\today}

\begin{abstract}

We study the entanglement dynamics for a two-spin system coupled
to a spin environment of different configurations by z-x type
interaction. The models considered in this paper are solved  both analytically and
numerically giving rise to some concise
analytical expressions when certain  approximations are properly made.
 Our purpose is to find how the initial states of the environment
with different numbers of spins affect the decay or revival of the entanglement between central qubits. In Particular, it is found that the block-entangled
 environment could speed
up the decay and revival of the qubit entanglement. Our results exhibit some interesting
features that have not been found for a boson bath.
\end{abstract}


\maketitle

\section{Introduction}

Entanglement of a quantum system not only has  important
consequences as a fundamental physical property, it has also been
identified as a resource in quantum information process (QIP),
quantum computing and quantum cryptography \cite{Nielson}.
Entanglement is fragile when the system of interest is not isolated
from the influence of its environments.  The dynamical aspect of
entangled state in the context of quantum open system has been
investigated in many different scenarios such as disentanglement of
qubit systems \cite{Zyczkowski,Yu-Eberly02,Abu2004,Yu-Eberly04PRL,review},
continuous variable systems \cite{Halliwell04,Paz08},  entanglement
delayed creation and revival \cite{Ficek07,Entbirth,Knight, Bose} and non-Markovian
entanglement evolution \cite{non-Markovian1,Hu2009}, to name a few.

As is well known, environmental noises that cause quantum systems to
decohere also lead to the loss of entanglement \cite{Paz2001,Zoller,Breuer-Petruccione}.
 In its simplest form the system plus environment can be examined in different
configurations of system-environment  models including  (i) harmonic
oscillators coupled to a bath of bosons, i.e. Quantum Brownian
motion model \cite{Hupazhang,zhangweimin}; (ii) Two-level atoms or spin-$1/2$ particles
(Qubits) interacting with a bath of harmonic oscillators, i.e.
spin-boson model \cite{Leggett, Schlosshauer-Hines-Milburn} and
(iii) Qubits  (or harmonic oscillators) coupled to a bath of spins-$1/2$, i.e. spin bath model
\cite{Shao-Ge,Yuan-Goan-Zhu,Goan1,Jing,Rao,Cecilia-Juan,
Hamdouni-Petruccione,Burkard,Ali-Chen-Goan,Privman1,Privman2,Privman3}.
It is shown that open system dynamics can be significantly modified by environmental configurations.

In this paper, we discuss entanglement dephasing dynamics for spin environment
models.  Our strategy is to directly solve the models without any approximations that
may lead to the conversion of bath spins into the effective Holstein-Primakoff bosons
or Schwinger bosons (e.g., see \cite{Yuan-Goan-Zhu,Jing}).
The purpose of this paper is to find important features induced from the spin environment
that are not seen in the boson case.   We not only consider
different couplings between the system and  its environments,  but also take into account
the effect of the various initial states of the environment on the entanglement dynamics of
the central qubit system.  Here we take our central qubit system as two uncoupled spin-1/2
particles (qubits), which are interacting with a bath of spins-1/2 particles
through z-x type coupling.  We have presented several analytical results on the dynamics
 of the two central qubits in physically relevant situations including individual coupling model and collective
coupling model.

This paper is organized as follows. In Sec.\ref{model} we present
the Hamiltonians of  two-qubit dephasing models with local and
global spin environments.  The analytical solutions to the both models are
provided. The entanglement dynamics of the central qubits with various initial states
are analytically investigated in Sec.\ref{local} for the local coupling, and in
Sec.\ref{global} for the global coupling,  respectively. Specifically,  for each model we consider
three different types of initial states of the environment: pure, mixed and
entangled.  In addition, we have extended our analytical treatments to numerical simulations for
more generic situations in Sec.\ref{comparison}, it is shown that the concurrence dynamics of the
central qubits is sensitively dependent on the total numbers of environment spins and bath initial
states. We conclude in Sec.\ref{conclusion}.

\section{Models and solutions}
\label{model}

\begin{figure}[!t]
\includegraphics[width = 12cm]{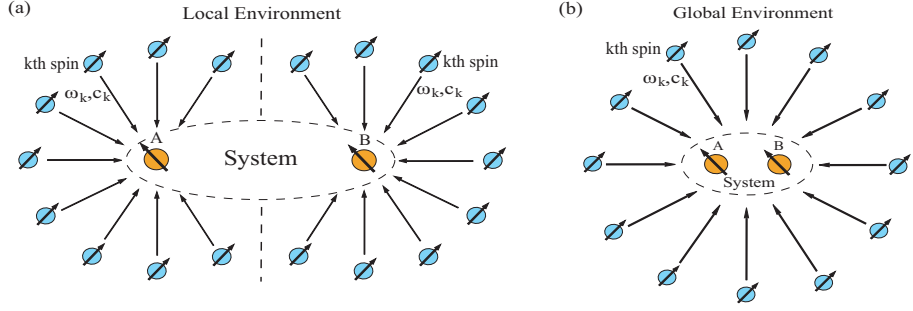}
\caption{(a) Two qubits are interacting with their local environments.
Two central qubits are prepared in an initially entangled state but have no
interaction between each other. (b) Two qubits are interacting with one
global environment.}\label{2model}
\end{figure}

 We consider two well-known cases for system and environment coupling:
Case~(a), the two system qubits are coupled to a common environment;
Case~(b), the two system qubits are interacting only with their local environments
and there is no crosstalk between the two environments.
For simplicity, the interaction between spins is neglected.
Precisely, the total Hamiltonian for these two cases are given by: \beqa \label{Hamiltonian}
H_{\rm{tot}}&=&H_{\rm{sys}}+H_{\rm{env}}+H_{\rm{int}},\\
H_{\rm{sys}}&=&\omega_A\si_z^A+\omega_B\si_z^B, \eeqa
where for the global environment case  we have $H_{\rm{env}}=\sum_{k}\omega_k\si_z^k$,
$H_{\rm{int}}=(\si_z^A+\si_z^B)\sum_kc_k\si_x^k$.  In the case of local
environment we have
$H_{\rm{env}}=\sum_{k}\omega_k\si_z^k+\sum_{l}\omega_l\si_z^l$,
$H_{\rm{int}}=\si_z^A\sum_kc_k\si_x^k+\si_z^B\sum_lc_l\si_x^l$.
$\omega_A$ and $\omega_B$ are the transition frequencies for the two
qubits, respectively. For simplicity, but without loss of generality, we
assume that $\omega_A=\omega_B=\omega_s$. Here, $\si_z$ and $\si_x$ are the
Pauli matrices, and $\omega_k$ is the frequency for $k$th spin in
the environment and $c_k$ are dimensionless real-value coupling
constants  between the $k$th spin and the central qubits. As is shown in
Fig.~\ref{2model}(a), $k$ and $l$ represent the two local baths for
the central qubits respectively. In both cases, the dephasing coupling ensures
that quantum coherence is damped, but the energy of central qubits is perfectly preserved.

For both cases, the dynamics for the central qubit systems can be solved analytically under the assumption that initially the system and environment are in a separable state, i.e. the density matrix for
the system plus environment at $t=0$ is of the following form:
\beq\rho(0)=\rho_{\rm{sys}}(0)\otimes\rho_B(0).
\eeq
The environment spins may take a simple form $\rho_B(0)=\prod_k\rho_B^k(0)$.  Otherwise, the environment may contain some entangled spin blocks, in this case, the environment state may be written as $\rho_B(0)=\prod_j\rho_B^j(0)$, where $\rho_B^j$ stands for the density matrix for each block (More precise descriptions are given in Subsection \ref{localentangled}).

For our dephasing model, the time-dependent density matrix for the total system may be written as:
\beq
\rho(t)= e^{-iH_{\rm{tot}}t}\rho(0)e^{iH_{\rm{tot}}t}
=e^{-iH_{\rm{sys}}t}[\prod_ke^{-i\hat{X}_kt}\rho(0)\prod_ke^{i\hat{X}_kt}]e^{iH_{\rm{sys}}t},
\eeq
where $H_{\rm{sys}}$ is the system Hamiltonian and $\hat{X}_k$ is the sum
of the bath Hamiltonian and interaction Hamiltonian of the
$k$th environment spin. For example, in case (b),
$\hat{X}_k=\omega_k\si_z^k+(\si_z^A+\si_z^B)c_k\si_x^k$. The
elements of the reduced density matrix (RDM) for the central qubit
system can be obtained by tracing over the bath spin degree of freedom:
\beqa \label{RDM} \rho_{mn}^{sys}(t)=\la m|\text{Tr}_k\rho(t)|n\ra
&=&e^{-i(E_m^\pr-E_n^\pr)t}\rho_{mn}^{\rm{sys}}(0)\prod_k\text{Tr}_kF_{mn}^k
=e^{-i(E_m^\pr-E_n^\pr)t}\rho_{mn}^{\rm{sys}}(0)f_{mn}(t),\\
f_{mn}(t)&=&\prod_k\text{Tr}_kF_{mn}^k=\prod_k\text{Tr}_k[e^{-i\hat{X}_k(E_m)t}\rho^k(0)e^{i\hat{X}_k(E_n)t}]. \label{ft}\eeqa
$|m\ra$, $|n\ra$ and $E_m^\pr$, $E_n^\pr$ are the eigenvectors and eigenvalues of the system Hamiltonian $H_{\rm{sys}}$ respectively. Note that the notations  $E_m, E_n$  denote the eigenvalues of the
coupling operator $\si_z^A+\si_z^B$,  thus $E_{m,n}^\pr=\omega_sE_{m,n}$. It should be noticed that the dephasing coefficient $f_{mn}(t)$ is a time dependent function satisfying $|f_{mn}(t)|\leq1$ and  $f_{nn}=1$.  The specific form of $f_{mn}(t)$ depends on the interaction Hamiltonian and the initial state of environment, which will be discussed in later sections.

\section{Solution of local coupling model with different initial states}
\label{local}

\subsection{General Solutions}

When each central qubit is interacting with its own environment, as shown in Fig.~\ref{2model}(a), the RDM of the two central qubits can be obtained by using Kraus operator technique \cite{Yu-Eberly04PRL,Yu-Eberly03}. In principle, we can always write the time-dependent RDM in such a way:
\beq
\rho(t)=\sum_{i}K_i\rho(0)K_i^\da, \eeq
where $K_i$  satisfy the condition $\sum_iK_i^\da K_i=I$. It is convenient to solve the model in this way since we may just focus on the subsystem involving only one qubit plus its local environment. The RDM of such a subsystem is thus a $2\times2$ matrix, which can be derived directly from Eq.~(\ref{RDM}):
\beqa \label{St1q}
\rho^A(t)&=&\left(\begin{array}{cc} \rho_{11}^A(0) & \rho_{12}^A(0)e^{-i2\omega_At}f(t) \\
\rho_{21}^A(0)e^{i2\omega_At}f(t)^* & \rho^A_{22}(0) \end{array} \right).
\eeqa
Obviously for the dephasing model, the diagonal elements are time independent, i.e. $\rho_{nn}(t)=\rho_{nn}(0)$, so we only need to consider the off-diagonal ones, for which $E_mE_n=-1$. Since $\rho_{12}=\rho^*_{21}$ for a $2\times2$ density matrix, we define $f(t)=f_{12}(t)$. The Kraus operators of one-qubit subsystem are given by:
\beqa
K_{1}^A=\left(\begin{array}{cc} 1 & 0 \\
0 & e^{-i2\omega_At}f(t)^* \end{array} \right),\ K_2^A=\left(\begin{array}{cc} 0 & 0 \\
0 & \sqrt{1-|f(t)|^2}\non\\ \end{array} \right),\\ \eeqa
where $f(t)$ is the dephasing coefficient defined in Eq.~(\ref{RDM}). Then the RDM for two qubit system can be directly obtained by:
\beq
\rho^{AB}(t)=\sum_{i,j=1,2}K_{ij}^{AB}\rho^{AB}(0)K_{ij}^{AB\da}, \eeq
where the composite Kraus operators  $K_{ij}=K_i^A\otimes K_j^B$. are given by
\beqa
K_{11} &=& \left(\begin{array}{cccc} 1 & 0 & 0 & 0\\
0 & e^{-it}f(t)^* & 0 & 0 \\0 & 0 & e^{-i2\omega_At}f(t)^* & 0 \\
0 & 0 & 0 & e^{-i4\omega_At}f(t)^{*2} \end{array} \right),\ \ \ \ \ \ \ \
K_{12} = \left(\begin{array}{cccc} 0 & 0 & 0 & 0\\
0 & \sqrt{1-|f(t)|^2} & 0 & 0 \\0 & 0 & 0 & 0 \\
0 & 0 & 0 & e^{-i2\omega_At}f(t)^*\sqrt{1-|f(t)|^2} \end{array} \right),\non\\
K_{21} &=& \left(\begin{array}{cccc} \ \ 0\ \ & \ \ 0\ \ & 0 & 0\\
0 & 0 & 0 & 0 \\0 & 0 & \sqrt{1-|f(t)|^2} & 0 \\
0 & 0 & 0 & e^{-i2\omega_At}f\sqrt{1-|f(t)|^2} \end{array} \right),\ \
K_{22}= \left(\begin{array}{cccc} 0\ \  & 0\ \  & 0\ \  & 0\\
0 & 0 & 0 & 0 \\0 & 0 & 0 & 0 \\
0 & 0 & 0 & 1-|f(t)|^2 \end{array} \right), \nonumber \eeqa
Then for an arbitrary initial state,the time dependent RDM is simply given by:
\beqa
\label{sol1}
\rho(t)=\left(\begin{array}{cccc} \rho_{11} & e^{-i2\omega_At}f(t)\rho_{12}
& e^{-i2\omega_At}f(t)\rho_{13} & e^{-i4\omega_At}f(t)^2\rho_{14} \\
e^{i2\omega_At}f(t)^*\rho_{21} & \rho_{22} & f(t)^2\rho_{23} & e^{-i2\omega_At}f(t)^*\rho_{24}\\
e^{i2\omega_At}f(t)^*\rho_{31} & f(t)^{*2}\rho_{32} & \rho_{33} & e^{-i2\omega_At}f(t)^*\rho_{34}\\
e^{i4\omega_At}f(t)^{*2}\rho_{41} & e^{i2\omega_At}f(t)\rho_{42} & e^{i2\omega_At}f(t)\rho_{43}
& \rho_{44}\end{array}\right).\non\\
\eeqa
Note that the solution (\ref{sol1}) is obtained without any approximations. We will discuss three special cases where the function $f(t)$ can be explicitly evaluated.


\subsection{Mixed State Environment}
\label{localmixed}

As our first example, we assume that the environment spins are in the following mixed state:
\beq \label{rhoT}
\rho^k(0)=\left(\begin{array}{cc} N_+^k & 0 \\ 0 & N_-^k
\end{array}\right),\eeq
where $N_+^k,N_-^k$ are the probability of which the $k$th spin is in up or down state ($N_+^k +N_-^k=1$). Here, we may choose $N_\pm^k=e^{\mp\beta\omega_k}/(e^{\beta\omega_k}+e^{-\beta\omega_k})$, where $\beta=1/T$ and $T$ is the temperature. By using the following identity
\beq \label{ezx}
e^{i(u\si_z+v\si_x)}=\cos\sqrt{u^2+v^2}+\frac{i\sin{\sqrt{u^2+v^2}}}{\sqrt{u^2+v^2}}
(u\si_z+v\si_x)
\non\eeq
and Eq.~(\ref{ft}), it is easy to show that:
\beq
\text{Tr}_k[F_{mn}^k]=\cos ^2(p_kt)+\frac{\sin^2(p_kt)}{p_k^2}(\omega_k^2+E_mE_nc_k^2)(N^k_++N^k_-)
=1-c_k^2(1-E_mE_n)\frac{\sin^2(p_kt)}{p_k^2},
\label{temp}
\eeq
where $p_k=\sqrt{\omega_k^2+c_k^2}$.\\[-5pt]

From (\ref{temp}),  we obtain:
\beqa \label{Smn}
\rho^A_{12}(t)&=&\rho_{12}^A(0)e^{-i(E_m^\pr-E_n^\pr)t}f(t),
\\ \label{ftdelta}
f(t)&=&\prod_k[1-2c_k^2\frac{\sin^2(p_kt)}{p_k^2}]. \eeqa
Eq.~(\ref{ftdelta}) is the exact solution in the condition of finite number of environment spins. It is easy to see that each environment spin contributes an oscillating function, which varies from $-1$ to $1$, affecting the coherence property of the central qubit system.  When the environment involves more and more spins, the overall effect of all spins results in a typically fast decaying function $f(t)$. It is evident from Eq.~(\ref{temp}) that the RDM is insensitive to the distribution of $N_+^k, N_-^K$. However, it is worth noting that this is not a generic feature for the spin environment. Actually, as shown in subsection \ref{commbath}, the coherence function $f(t)$ can explicitly depend on the distributions $N_+^k, N_-^K$.

Interestingly, if more information about the bath spectrum density becomes available, we may get a rather
compact form for the function $f(t)$. For example, for a bath with infinite numbers of spins and the Ohmic spectrum density,
\beq
J(\omega)=\eta\frac{\omega}{c^2(\omega)}e^{-\frac{\omega}{\omega_c}},
\eeq
with the weak coupling approximation $c_k\ll1$, we can get:
\beq \label{ftOhmic} f(t) =
(1+4\omega_c^2t^2)^{-\frac{1}{2}\eta}. \eeq
The power law decay of quantum coherence clearly deviates from the exponential decay commonly seen in the case of Markov bosonic bath \cite{Zoller}.

\subsection{Pure State Environment}
\label{localpure}

Another interesting case is that the bath spins are initially prepared in a pure state:
$$|\Psi(0)\rangle=\cos\alpha_k|\uparrow\rangle+\sin\alpha_k|\downarrow\rangle.$$
The density matrix may be written as,
\beq \label{pure} \rho^k(0)=|\Psi(0)\rangle\langle\Psi(0)|=\left(\begin{array}{cc} \cos^2\alpha_k &
\cos\alpha_k\sin\alpha_k \\ \cos\alpha_k\sin\alpha_k &
\sin^2\alpha_k
\end{array}\right). \eeq
The RDM for the central qubits at time $t$ is also determined by the same terms as in
Eq.~(\ref{RDM}), thus we have
\beq
\text{Tr}_k[F_{mn}^k]=1-\frac{c_k^2}{p_k^2}\sin^2(p_kt)(1-E_mE_n)+i\frac{c_k}{2p_k}\sin(p_kt)\sin2\alpha_k(E_n-E_m). \label{pureinitial}\eeq
Eq.~(\ref{pureinitial}) is different from Eq.~(\ref{temp}) by an imaginary factor in the order
of $c_k/p_k$. Finally, the time dependent dephasing coefficient is given by
\beq
f(t)=\prod_k[1-2\frac{c_k^2}{p_k^2}\sin^2(p_kt)+i\frac{c_k}{p_k}\sin(p_kt)\sin2\alpha_k].
\eeq
We note that the above expression does not permit $f(t)$ to have the same compact and useful form as Eq.~(\ref{ftOhmic}) for the continuous spectrum density considered before.

\subsection{Environment with Entangled Blocks}
\label{localentangled}

We also consider a situation where the environment spins may be decomposed into a set of entangled blocks. This clearly is an interesting configuration that may impose some more strong constraints on the central qubit entanglement evolution. More specifically, we begin with an environment consisting of $N\times M$ spins in total. We may divide the total $N\times M$ spins into $M$ blocks,  each contains exactly $N$ spins, which is prepared in an $N$-GHZ state \cite{GHZ}. In this case, the matrix elements of the RDM are given by Eq.~(\ref{RDM}) with a modified dephasing function $f_{mn}(t)$:
\beqa \non
f_{mn}(t)=\prod_{j=1}^M\text{Tr}^N_k\big[\prod_{k=1}^Ne^{-i\hat{X}_kt}\rho_B^{N,j}(0)\
\prod_{k=1}^Ne^{i\hat{X}_kt}\big]=\prod_{j=1}^M\text{Tr}^N_k[F_{mn}^{N,j}(t)]
\eeqa
where $\rho_B^{N,j}(0)$ is the density matrix for $N$ entangled bath spins confined in
the $j$th block. And we have
\beqa
\text{Tr}F_{mn}^{N,j}(t)&=&\prod_{k=1}^N(1-\frac{c_k^2}{p_k^2}\sin^2p_kt(1-E_mE_n))
+(\frac{E_n-E_m}{2})^N\prod_{k=1}^N(\frac{1-\cos2p_kt}{p_k^2}\omega_kc_k+i\frac{\sin2p_kt}{p_k}c_k)\non\\
& &+(\frac{E_n-E_m}{2})^N\prod_{k=1}^N(-\frac{1-\cos2p_kt}{p_k^2}\omega_kc_k+i\frac{\sin2p_kt}{p_k}c_k). \eeqa

A particular expression for the coherence function $f(t)$ may be obtained when each block only contains
two entangled spins, which are prepared
as $\Phi_j^{1,2}=\frac{1}{\sqrt2}(|\downarrow\downarrow\rangle+|\uparrow\uparrow\rangle)$
and $c_{j1}=c_{j2}=c_j,p_{j1}=p_{j2}=p_j$, then the off-diagonal terms in the RDM are:
\beqa
\rho^S_{m\neq n}(t) &=& e^{-i(E_m-E_n)t}\rho^S_{m\neq n}(0)f(t),\\
f(t)&=&\prod_{j=1}^M (1-2\frac{c_j^2}{p_j^2}\sin^2 2p_jt). \eeqa

Again, with the Ohmic spectrum density assumption and $c_k<<1$,the dephasing coefficient $f(t)$ in the long time limit is simply given by:
\beq \label{ohmic} f(t)=
(1+16\omega_c^2t^2)^{-\frac{1}{2}\eta}\approx(4\omega_ct)^{-\eta}. \eeq
Immediately, we can see from the above calculations that the block-entangled environment may give rise to more severe decoherence effect on the central qubits than the mixed initial state (e.g., see Eq.~(\ref{ftOhmic})). It is interesting to consider another simple case:
$c_{j1}=-c_{j2}=c_j,p_{j1}=p_{j2}=p_j$, with which we get
\beq f(t)=\prod_{j=1}^M
(1-8\frac{\omega_j^2c_j^2}{p_j^4}\sin^4p_jt). \eeq

Using the same assumption and approximation above, we get a simple function in the long time limit:
\beq
f(t)=(1+16\omega_c^2t^2)^{\frac{1}{2}\eta}(1+4\omega_c^2t^2)^{-2\eta}\approx(4\omega_c^2t^3)^{-\eta}.
\eeq

\section{Solution of global coupling model with different initial states}
\label{global}

\subsection{General Solution}
In this subsection, we consider the case that both qubits are coupled to the same environment, illustrated in Fig.~\ref{2model}(b). Again, this model can be solved exactly by using the same method as we used in the previous section, but with different conventions $E_m,E_n\in\{2,0,0,-2\}$. It can be seen below that new features can arise from the backreaction on the central qubits induced by the common bath. Thus, the resultant RDM is different and more complex, which is given by:
\beq \label{rt}
\left(\begin{array}{cccc} \rho_{11} & \rho_{12}e^{-i2\omega_st}g(t) &
\rho_{13}e^{-i2\omega_st}g(t) &
\rho_{14}e^{-i4\omega_st}f(t) \\
\rho_{21}e^{i2\omega_st}g(t)^* & \rho_{22} & \rho_{23} & \rho_{24}e^{-i2\omega_s t}g(t)^*\\
\rho_{31}e^{i2\omega_st}g(t)^* & \rho_{32} & \rho_{33} & \rho_{34}e^{-i2\omega_st}g(t)^* \\
\rho_{41}e^{i4\omega_st}f(t)^* & \rho_{42}e^{i2\omega_st}g(t) &
\rho_{43}e^{i2\omega_st}g(t) & \rho_{44} \end{array}\right), \eeq
for simplicity. Considering the the initial states of system qubits $A,B$ are in Bell states. When
$\Psi_{AB}=\frac{1}{\sqrt2}(|\uparrow\uparrow\rangle\pm|\downarrow\downarrow\rangle)$,
the result RDM will be similar as in Eq.~(\ref{sol1}) with $f(t)^2\rightarrow f(t)$. Yet for the other two states $\Psi_{AB}=\frac{1}{\sqrt2}(|\uparrow\downarrow\rangle\pm|\downarrow\uparrow\rangle)$,
it can be shown that entanglement is perfectly preserved.

\subsection{Mixed State Environment}
\label{commbath}

For the environment spins that are initially in a mixed state, we have
\beqa f(t)&=&\prod_k[1-8\frac{c_k^2}{q_k^{2}}\sin^2(q_k t)], \\\non
g(t)&=&\prod_k\left\{\cos(q_kt)\cos(\omega_kt)+\omega_k\frac{\sin(q_k
t)\sin(\omega_k t)}{q_k}+i\Delta_k\left[\cos(q_kt)\sin(\omega_k t)
-\frac{\omega_k}{q_k}\sin(q_k t)\cos(\omega_kt)\right]\right\}.\eeqa
where $q_k=\sqrt{\omega_k^2+4c_k^2}$, $\Delta_k=\frac{N_+^k-N_-^k}{N_+^k+N_-^k}$.

\subsection{Pure State Environment}

If the bath is initially in a pure state (see Eq.~(\ref{pure}), by assuming $\alpha_k=\pi/4$), we may get an explicit form of the density matrix Eq.~(\ref{rt}) with
\beqa  f(t)&=&\prod_k\{ 1-8\frac{c_k^2}{q_k^{2}}\sin^2(q_kt)+
i2\frac{c_k}{q_k}\sin[2q_kt]\},\\ \non
g(t)&=&\prod_k\{\cos(q_kt)\cos(\omega_kt)+\frac{\omega_k}{q_k}\sin(q_kt)\sin(\omega_kt)
-i\frac{2c_k}{q_k}\sin(q_kt)\cos(\omega_k t)\}. \eeqa
This solution will be used for entanglement dynamics to be discussed later.\\

\subsection{Environment with Entangled Blocks}

For simplicity, we consider a case in which the environment has $M$ blocks, each block consists of two entangled spins (i.e., $N=2$). For this structured environment, the time-dependent coefficients in Eq.~(\ref{rt}) takes the following form:
\beqa\label{gcoefmm}
f&=&\prod_j^M[(1-8c_{j,1}^2S_{j,1}^2)(1-8c_{j,2}^2S_{j,2}^2) +16c_{j,1}c_{j,2}(\omega_{j,1}\omega_{j,2}S_{j,1}^2S_{j,2}^2-C_{j,1}C_{j,2}S_{j,1}S_{j,2})],\\
g&=&\prod_j^M[(C_{j,10}C_{j,1}+S_{j,10}S_{j,1}\omega_{j,1}^2)(C_{j,20}C_{j,2}+S_{j,20}S_{j,2}\omega_{j,2}^2)\non\\
&-&\omega_{j,1}\omega_{j,2}(C_{j,10}S_{j,1}-S_{j,10}C_{j,1})(C_{j,20}S_{j,2}-S_{j,20}C_{j,2})
-4c_{j,1}c_{j,2}S_{j,1}S_{j,2}\cos(\omega_{j,1}+\omega_{j,2})].
\eeqa
with the notation
\beqa \non
C_{j,i}=\cos(q_{j,i}t), S_{j,i}=\frac{\sin(q_{j,i} t)}{q_{j,i}},
C_{j,i0}=\cos(\omega_{j,i}t), S_{j,i0}=\sin(\omega_{j,i}t),
q_{j,i}=\sqrt{\omega_{j,i}^2+4c_{j,i}^2}, \ i=1,2.
\eeqa
where $\omega_{j,i}$ and $c_{j,i}$ are the frequency and coupling strength for the $i$th spin in the $j$th block.

\section{Entanglement Dynamics For Spin Environments}
\label{comparison}

In this paper we use Wootters' concurrence to measure the entanglement of the central qubits \cite{Wootters}, which is defined as
\beq
C(\rho)=\max\{0,\sqrt\lambda_1-\sqrt\lambda_2-\sqrt\lambda_3-\sqrt\lambda_4\},
\eeq
where $\lambda_i$ are the eigenvalues in decreasing order of the matrix
\[\zeta=\rho(\si_y\otimes\si_y)\rho^*(\si_y\otimes\si_y).\]
For our model all the off-diagonal elements in the RDM are obtained by their initial value times a
dephasing coefficient, thus if any element is zero at $t=0$, then it will remain zero for any $t>0$. Consequently, when Bell states are chosen as the initial state for the central qubit system, its RDM will preserve an X-form \cite{Yu-Eberly03}. In this case, the concurrence may be simplified as:
\beqa \label{con}
C(\rho)=2\max\{0,|\rho_{14}(t)|-\sqrt{\rho_{22}(t)\rho_{33}(t)},|\rho_{23}(t)|-\sqrt{\rho_{11}(t)\rho_{44}(t)}\}
=2\max\{0,|f^2(t)|\}. \eeqa

It is clear that the local environments do not generate entanglement between central qubits, so our focus will be on entanglement decay.  For the global environment, however, it is possible to achieve
entanglement generation when initially the central qubits are in a separable state. Based on our analytical solutions to both models, we will be able to systematically study concurrence dynamics $C$ of the central qubits in the following two subsection. In all the numerical results presented below, we will make the following conventions: the frequency of system qubits is chosen as $\omega_s=1$; for the environment spins, their frequencies and coupling constants are randomly distributed in the intervals $\omega_k\in(1, 2)$,  $c_k\in(0.1,0.2)$, respectively.

\subsection{Entanglement Decay in Local Environment Model}

\begin{figure}[!t]
\includegraphics[width = 8cm]{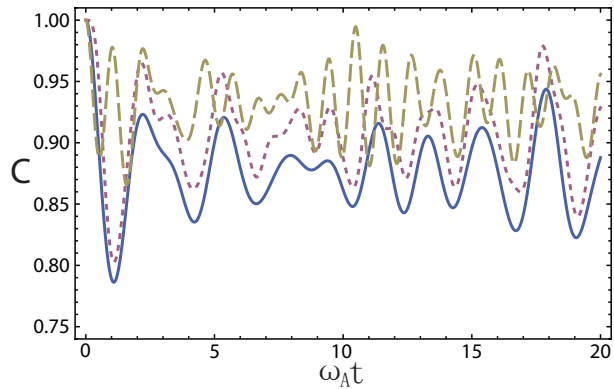}
\caption{Two qubits interacting locally with two 6-spin environments. The blue solid line is for environment initially in mixed state, red doted line for pure state, and yellow dashed line for block-entangled environment.}\label{sep6}
\end{figure}

\begin{figure}[!t]
\includegraphics[width = 8cm]{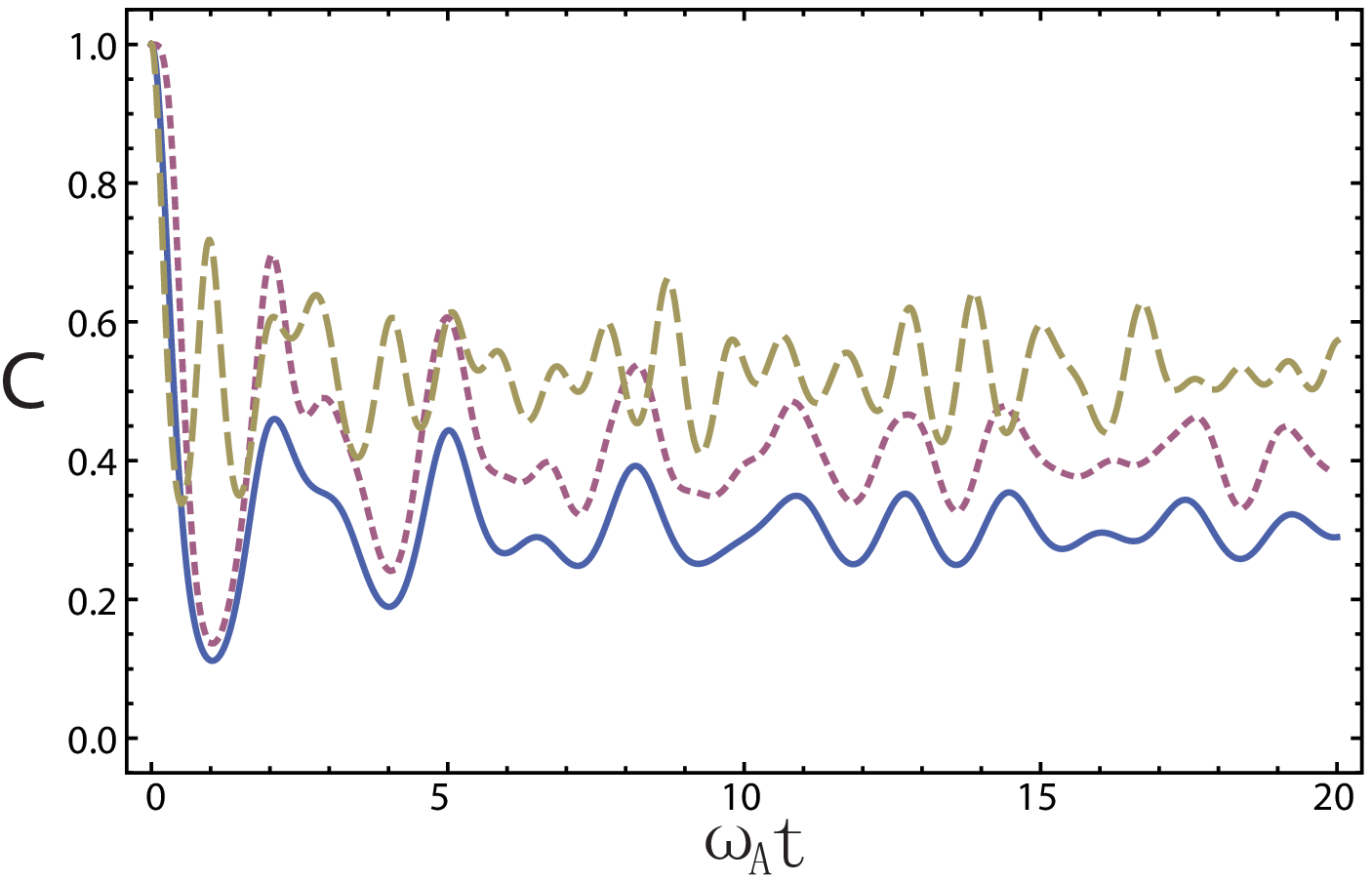}
\caption{Two qubits interacting locally with two 60-spin environments.}\label{sep60}
\end{figure}

\begin{figure}[!t]
\includegraphics[width = 8cm]{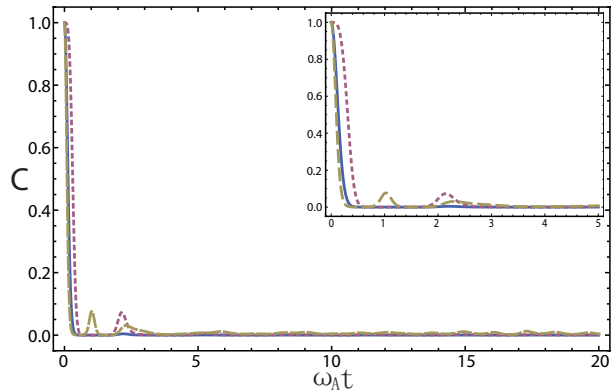}
\caption{Two qubits interacting locally with two 400-spin environments.}\label{sep400}
\end{figure}

Now we consider the central qubit system to be prepared in the Bell state $\Psi_{AB}=\frac{1}{\sqrt2}(|\uparrow\uparrow\rangle+|\downarrow\downarrow\rangle)$.
Figs.~\ref{sep6}, \ref{sep60} and \ref{sep400} show the concurrence dynamics with $6$, $60$ and $400$ environment spins. Here and next section, the blue solid line is always used for the situation that the environment is in the mixed state (for local environment $\Delta_k$ is not important,
but we have set $\Delta_k=0.9$ in next subsection for global environment.), the red dotted line is for environment in the pure state and the yellow dashed one is for block-entangled environment.

If the number of the environment spins is not too large, as illustrated in Figs.~\ref{sep6} and \ref{sep60}, the central qubit entanglement is a quasi-periodic function of $t$, representing a strong non-Markovian feature of environment since the information transferred to the environment from the system will quickly go back to itself. It should be noted that block-entangled environment could play a positive role in preserving the system concurrence compared with the cases of mixed or pure initial states. In general, the mixed environment state may cause more detrimental effect to entanglement of central qubits.

Typically, a larger environment shows more dramatic differences between the three initial states. For the $400$-spin bath shown in Fig.~\ref{sep400}, we see that the initial entanglement decays to zero much quicker than the case of small environments. This is consistent with our general observations
for the bosonic case. In all the three cases, the increased degrees of freedom of the environment will cause effectively an irreversible information flow between the system and its environment. It is noticed that concurrence revival is found in the case of both the block-entangled environment and pure state,  but not in the mixed state case. For our example, as shown in Figs.~\ref{sep6} and \ref{sep60}, entanglement decay and revival occur approximately at the same pace for mixed and pure initial states. But for the block-entangled bath, the structured environment may speed up both decay and revival processes.

\subsection{Entanglement Generation in Global Environment Model}

\begin{figure}[!t]
\includegraphics[width = 8cm]{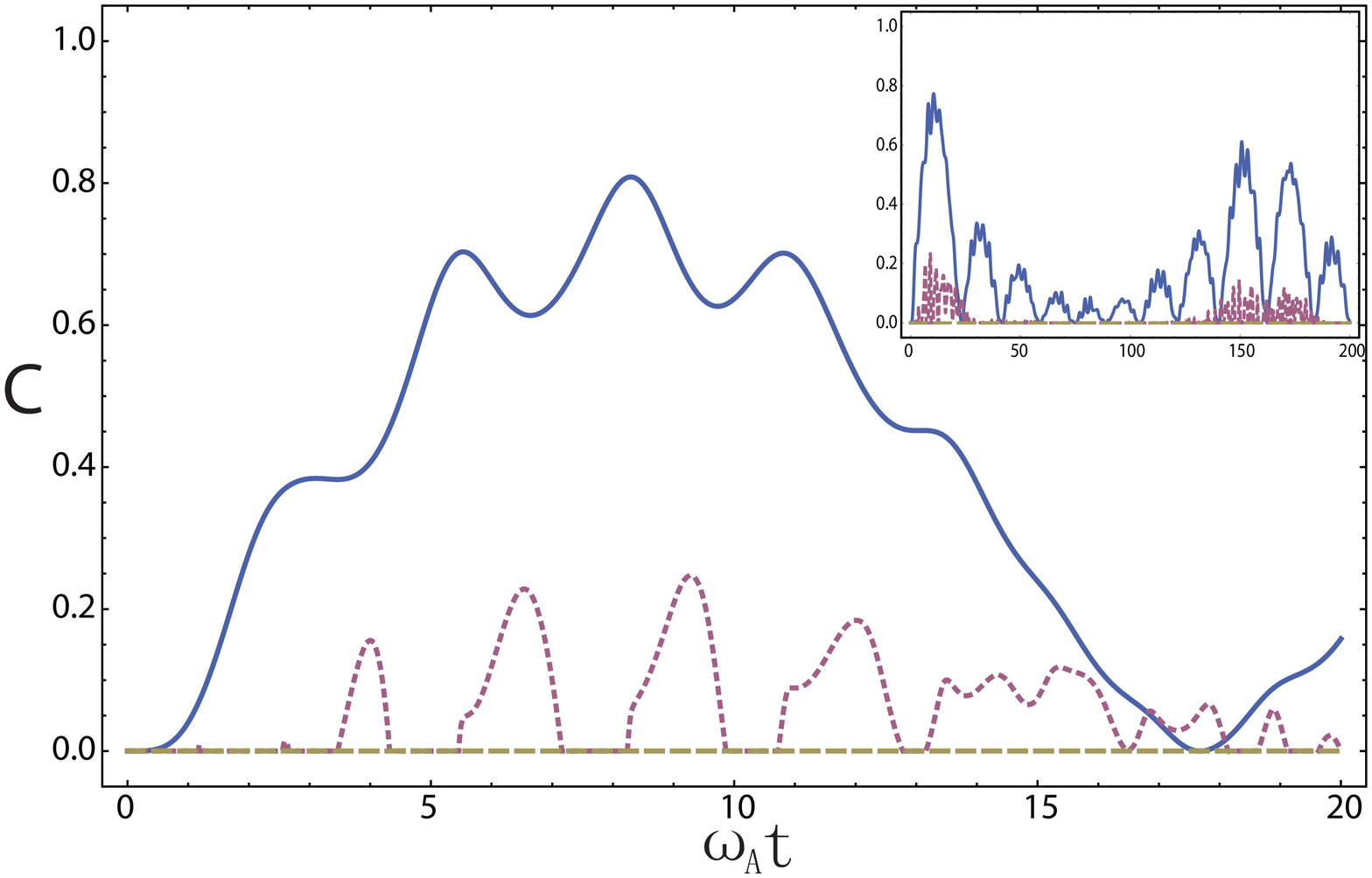}
\caption{Two qubits interacting with a 6-spin global environment.
The blue solid line is for environment initially in mixed state, red dotted line is for pure state,
yellow dashed line is for pair-wise entangled bath.}\label{comm6}
\end{figure}

\begin{figure}[!t]
\includegraphics[width = 8cm]{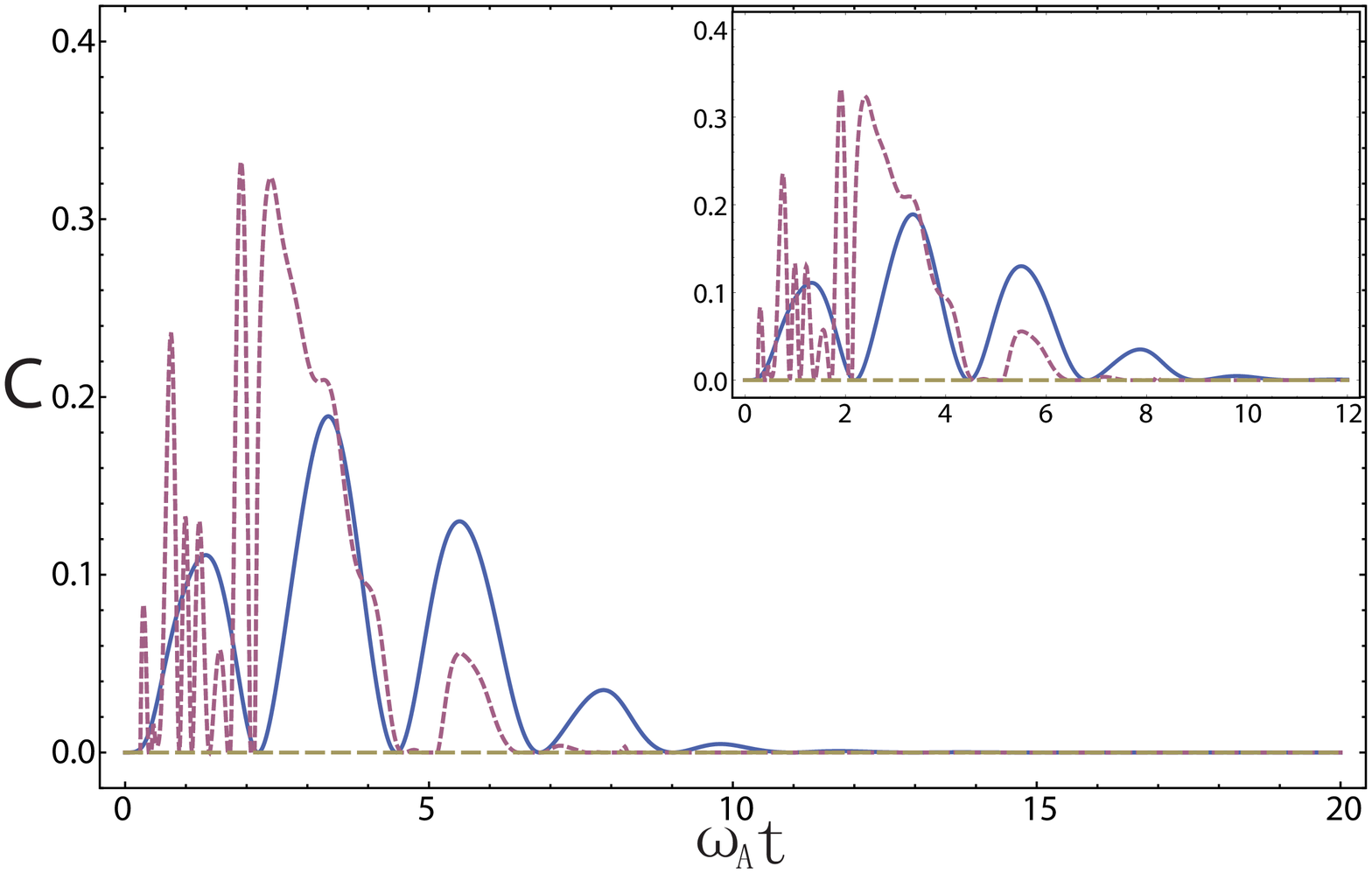}
\caption{Two qubits interacting with a 60-spin global environment.}\label{comm60}
\end{figure}

\begin{figure}[!t]
\includegraphics[width = 8cm]{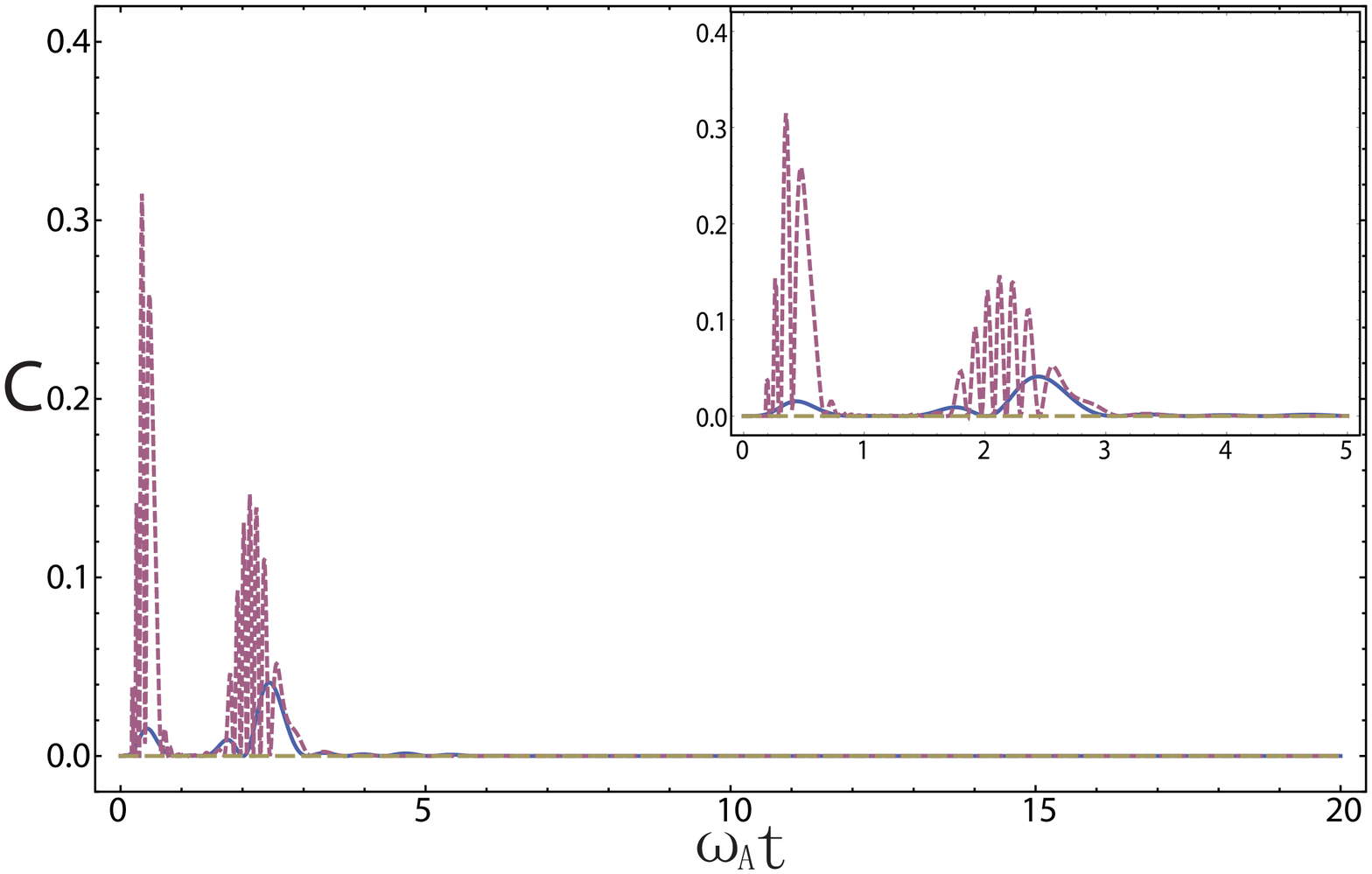}
\caption{Two qubits interacting with a 200-spin  global environment.}\label{comm200}
\end{figure}

\begin{figure}[!t]
\includegraphics[width = 8cm]{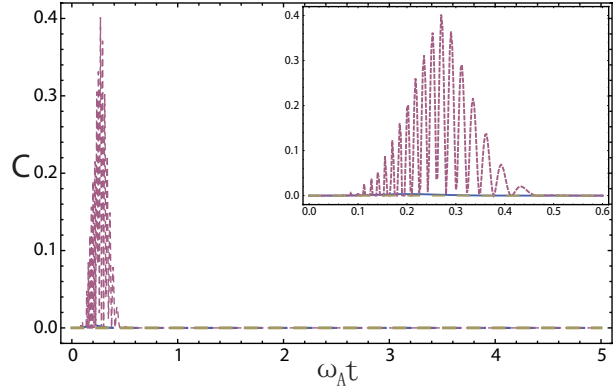}
\caption{Two qubits interacting with a 800-spin global environment.}\label{comm800}
\end{figure}

When the system qubits are coupled to a global environment, it is possible to achieve entanglement generation between the two central qubits. It is clear that such generation of entanglement is partly due to the back-reaction on the central qubits through system-environment couplings. For example, let us consider a separable initial state of the central qubits $A,B$ represented by
$\Psi_{AB}=\frac{1}{2}(|\uparrow\rangle_A+|\downarrow\rangle_A)\otimes(|\uparrow\rangle_B+|\downarrow\rangle_B)$.
In what follows we show that entanglement generation is sensitively dependent on the environment initial states.

We begin with the case of small environment with only 6 spins, as shown in Fig.~\ref{comm6}.
We see again a quasi-periodic evolution as expected for a small environment. A more interesting phenomenon is the generation of entanglement by the mixed state bath. We have found that the mixed initial state is very efficient in producing high degree of entanglement. Moreover, it shows that concurrence proceeds in time in a predictable manner for a long period of time. Although the pure state environment can also generate entanglement, but the degree is much lower than the mixed state. For the case of the block-entangled environment (it is assumed each pair is in a GHZ state), surprisingly we find that,  for possible configurations, there is no generation of entanglement.

If the environment contains a large number of spins, the pure state environment may generate a
high degree of entanglement. On the contrary, for the mixed state environment, less entanglement is
generated. Fig.~\ref{comm60} shows the entanglement dynamics for two central qubits coupled to a
global bath with $60$ spins. The entanglement generated may only survive some time before decaying
to zero, and stay as zero for a long period of time (at least no revival is found in our simulations).
However, the entanglement generated by the mixed state environment shows a regular creation and decay
pattern that is not seen for the pure initial state.
Similar oscillating patterns are also observed when the number of spins is increased to $200$.
In Fig.~\ref{comm200},  two bumps are found at early stage, and no visible entanglement is detected at a late stage. This phenomenon occurs when the environment is larger than the size of system,  but not large enough to trap all the information leaked into the environment from the system.

Finally, we plot Fig.~\ref{comm800} for a very large environment consisting of  $800$ spins. In many ways, the model under this condition may be well represented by a Markov system. Intuitively, for such a large environment, the system information once lost into the environment won't come back to itself again.  For our example, entanglement generation is not observed  for the mixed state bath. But for the pure state bath, we see $0.4$ entanglement generation measured by concurrence in an early stage due to the initial relaxation of the environment into its steady state. Such an initial generation of entanglement
is also observed in the Holstein-Primakoff boson case \cite{Yuan-Goan-Zhu, Jing}.

\section{Conclusion}
\label{conclusion}

We studied dynamics of entanglement of the open system under the z-x interaction
 with different spin environments. The all models presented here are
solved analytically.  The power law decay of entanglement is obtained
when the environment is large and the spectrum density is of an ohmic type.

For the local environment model, we found that, in the case of the small environment,
the central qubits can typically maintain a high level entanglement for a long period of time.
Interestingly, we have shown that the block-entangled environment provides stronger
protection of the entanglement. In addition, the entangled spins confined in the local environments
can accelerate the decay and revival of the entanglement between central qubits.

For the global environment model, we found that, in the case of small
environments (see Fig.~\ref{comm6}), the concurrence evolves in a similar
pattern when driven by a pure state or mixed state baths. Even for a large
environment (see, Fig.~\ref{comm60} and Fig.~\ref{comm200}), the entanglement
generated is still seen to decay in a non-monotonic way, a feature that is
commonly ascribed to the non-Markovian dynamics. For a very large environment
(see, Fig.~\ref{comm800}), similar to Markov limit, the mixed state environment
considered here does generate entanglement. However, the pure state bath
entanglement generation at early stage is possible. As for block-entangled bath
with GHZ state as the initial state, the generation of entanglement is always
suppressed in the examples considered here; a more detailed discussion of this
interesting phenomenon will be discussed in another paper.

\begin{acknowledgements}
We acknowledge grant support from DARPA QuEST HR0011-09-1-0008,
the NSF PHY-0925174, and the National Natural Science Foundation
of China (10804069).
\end{acknowledgements}

 \end{document}